\documentclass[aip,apl,reprint]{revtex4-1} 
\usepackage{graphicx}
\usepackage{dcolumn}
\usepackage{bm}
\usepackage{amsmath, amssymb}
\usepackage{color}

\begin{document}

\title{Domain wall-based spin-Hall nano-oscillators}

\author{N. Sato}
\affiliation{Helmholtz-Zentrum Dresden--Rossendorf, Institut f\"ur Ionenstrahlphysik und Materialforschung, D-01328 Dresden, Germany}

\author{K. Schultheiss}
\affiliation{Helmholtz-Zentrum Dresden--Rossendorf, Institut f\"ur Ionenstrahlphysik und Materialforschung, D-01328 Dresden, Germany}

\author{L. K\"orber}
\affiliation{Helmholtz-Zentrum Dresden--Rossendorf, Institut f\"ur Ionenstrahlphysik und Materialforschung, D-01328 Dresden, Germany}
\affiliation{Technische Universit\"at Dresden, 01062 Dresden, Germany}

\author{N. Puwenberg}
\affiliation{Leibniz Institute for Solid State and Materials Research (IFW) Dresden, 01069 Dresden, Germany}

\author{T. M\"uhl}
\affiliation{Leibniz Institute for Solid State and Materials Research (IFW) Dresden, 01069 Dresden, Germany}

\author{A.A. Awad}
\affiliation{Department of Physics, University of Gothenburg, 412 96 Gothenburg, Sweden}

\author{S.S.P.K. Arekapudi}
\affiliation{Institut f\"ur Physik, Technische Universit\"at Chemnitz, D-09107 Chemnitz}

\author{O. Hellwig}
\affiliation{Helmholtz-Zentrum Dresden--Rossendorf, Institut f\"ur Ionenstrahlphysik und Materialforschung, D-01328 Dresden, Germany}
\affiliation{Institut f\"ur Physik, Technische Universit\"at Chemnitz, D-09107 Chemnitz}

\author{J. Fassbender}
\affiliation{Helmholtz-Zentrum Dresden--Rossendorf, Institut f\"ur Ionenstrahlphysik und Materialforschung, D-01328 Dresden, Germany}
\affiliation{Technische Universit\"at Dresden, 01062 Dresden, Germany}

\author{H. Schultheiss}
\affiliation{Helmholtz-Zentrum Dresden--Rossendorf, Institut f\"ur Ionenstrahlphysik und Materialforschung, D-01328 Dresden, Germany}
\affiliation{Technische Universit\"at Dresden, 01062 Dresden, Germany}

\date{\today}


\maketitle

{\bf 
In the last decade, two revolutionary concepts in nano magnetism emerged  from research for storage technologies and advanced information processing. The first  suggests the use of magnetic domain walls (DWs) in ferromagnetic nanowires to permanently store information in DW racetrack memories \cite{Parkin2008}. The second proposes a hardware realisation of neuromorphic computing in nanomagnets using nonlinear magnetic oscillations in the GHz range \cite{Locatelli,Grollier2017}. Both ideas originate from the transfer of angular momentum from conduction electrons to localised spins in ferromagnets\cite{Slonczewski1996, Berger1996}, either to push data encoded in DWs along nanowires or to sustain magnetic oscillations in artificial neurones. Even though both concepts share a common ground, they live on very different time scales which rendered them incompatible so far. Here, we bridge both ideas by demonstrating the excitation of magnetic auto-oscillations inside nano-scale DWs using pure spin currents. }


The spin-tranfer-torque (STT) effect discovered in 1996 by Slonczewski and Berger \cite{Slonczewski1996, Berger1996} allows the manipulation of localised magnetic moments in a ferromagnet by the transfer of spin angular momentum from spin polarised conduction electrons. The  direction of the magnetisation can either be switched permanently\cite{Myers} or  can be forced to oscillate at radio frequencies \cite{Kiselev}. Quite soon thereafter it was recognised that a charge current in a ferromagnet, which is intrinsically spin polarised, can move magnetic domain walls (DWs) in nanowires \cite{CIDWM}. This  gave rise to the idea of the magnetic racetrack memory \cite{Parkin2008} and, quite recently, current induced skyrmion motion \cite{Skyrme,Muhlbauer,Fert2013}. While these schemes target nonvolatile, long term data storage, the STT effect in spin-torque nano-oscillators can be exploited to drive magnetic auto-oscillations\cite{Slavin} and, eventually, to radiate spin waves \cite{Madami} by compensating the intrinsic magnetic damping.  Another leap was the development of spin-Hall nano-oscillators (SHNO) \cite{Demidov2012,Demidov2014} 
in which pure spin currents are generated via the spin-Hall effect (SHE)\cite {Dyakonov1971,Hirsch1999,Ando2008, Hoffmann2013} and by which even propagating spin waves are excited\cite{Urazhdin2016,Divinskiy2018}. This puts SHNOs at the heart of magnonics \cite{Neusser2009,Kruglyak2010,Lenk2011} which proposes a novel type of low energy, non boolean computing based on magnons, the quanta of spin waves, as carriers of information\cite{Chumak2015} or even neuromorphic computing  \cite{Grollier2017} based on the  nonlinear character of magnonic auto-oscillations. In a previous work, we demonstrated that DWs can channel magnons in an effective magnetic potential well \cite{Wagner}. This raised the question if a magnetic DW can potentially be a self-organized, movable SHNO.
 
\begin{figure}[b]
\begin{center}
\scalebox{1}{\includegraphics[width=8cm]{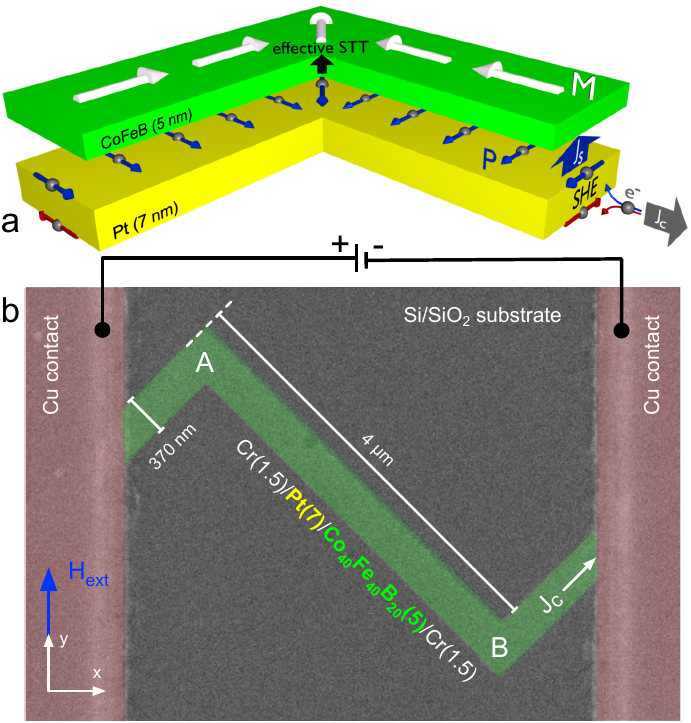}}
\caption{\label{fig1}\textbf{Sample layout and working principle.} {\bf a} Schematic  of a SHNO based on a domain wall. The functional layers of Pt and Co$_{40}$Fe$_{40}$B$_{20}$ are separated only for illustration purposes. {\bf b}, SEM image of the investigated 370-nm wide nanowire that exhibits two 90$^{\circ}$ bends at positions A and B and is connected to a current source by two Cu contacts.  
}
\end{center}
\end{figure}

In order for the STT effect to counteract the intrinsic magnetic damping, the magnetisation $\vec{\bf M}$ has to have a component antiparallel to the polarisation $\vec{\bf P}$ of the injected spin current $I_\mathrm{s}$ with density $\vec{\bf J}_\mathrm{s}$. Due to spin-dependent scattering associated with the SHE, a charge current $I_\mathrm{c}$ with density $\vec{\bf J}_\mathrm{c}$ flowing in a heavy metal is converted into a transverse pure spin current  with $\vec{\bf J}_\mathrm{s}  \bot  \vec{\bf J}_\mathrm{c}\bot \vec{\bf P}$ (Fig.~\ref{fig1}a). In a conventional nanowire, therefore, an external magnetic field is needed to overcome the shape anisotropy and to align $\vec{\bf M} \bot \vec{\bf J}_\mathrm{c}$ \cite{Duan}. Additionally, $I_\mathrm{s}$ needs to be high enough to exert sufficient torque and compensate the magnetic damping. Typically, this is achieved by lithographically patterning devices with active areas of only few $10-100$ nanometres\cite{Amad}.

 However, the sizes are limited by lithography and the position of the active area is fixed. To overcome these limitations, we investigated the excitation of auto-oscillations inside magnetic DWs which bring multiple benefits:  they can be created and destroyed dynamically, are very small, inherently exhibit $\vec{\bf M} \bot \vec{\bf J}_\mathrm{c}$ and can easily be moved by electric currents, lasers, and small magnetic fields. This Letter demonstrates the possibility to drive strongly localised magnon auto-oscillations inside 90$^\circ$  DWs before they are moved by current induced DW motion.

At the heart of any SHNO is a bilayer stack of a heavy metal in direct contact to a ferromagnet (Fig.~\ref{fig1}). In our experiments, we use Cr(1.5)/Pt(7)/Co$_{40}$Fe$_{40}$B$_{20}$(5)/Cr(1.5) multilayers, with Pt and CoFeB as the functional materials. Two Ta(5)/Cu(150)/Ta(5) contacts allow for the application of a direct charge current $I_\mathrm{c}$ with density $\vec{\bf J}_\mathrm{c}$. All thicknesses are given in nanometers. As is depicted in Fig.~\ref{fig1}a, the SHE in Pt converts $\vec{\bf J}_\mathrm{c}$ into a transverse pure spin current $I_\mathrm{s}$ with density $\vec{\bf J}_\mathrm{s}$ and polarity $\vec{\bf P}$. For a parallel (anti-parallel) orientation of $\vec{\bf P}$ and $\vec{\bf M}$, the STT leads to an increased (decreased) effective damping of the magnetisation dynamics in the CoFeB. To reproducibly control the formation of a DW, we pattern the functional stack to a 370-nm wide zig-zag shaped nanowire with two 90$^{\circ}$ bends, as highlighted in light green in the scanning electron microscopy (SEM) image in Fig.~\ref{fig1}b. In order to define the magnetic ground state, the sample can be saturated by an external magnetic field $\vec{\bf H}_\mathrm{ext}$ applied along the $y$-axis.  

\begin{figure}
	\begin{center}
		\scalebox{1}{\includegraphics[width=8.3cm]{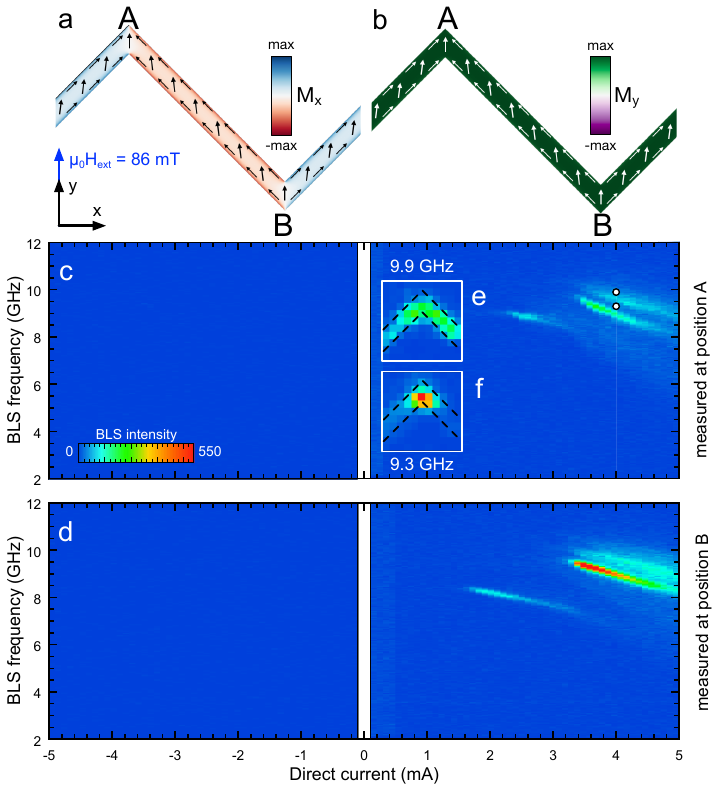}}
		\caption{\label{fig2}\textbf{Auto-oscillations in the nanowire with applied field.}~ Micromagnetic simulation of {\bf a}, the $x$- and {\bf b}, the $y$-component of the equilibrium magnetisation under an applied field of  $\mu_{0} \mathrm{H}_\mathrm{ext}=86$\,mT. {\bf c},{\bf d}, BLS spectra recorded at positions A  and B, respectively, with an external field of $\mu_{0} \mathrm{H}_\mathrm{ext}=86$\,mT covering the current range between $-0.1$ to $-5$\,mA and $0.1$ to $5$\,mA, respectively. Insets {\bf e},{\bf f} show the two-dimensional intensity distributions of auto-oscillations excited at 9.3 and 9.9\,GHz with $I_\mathrm{c}=4$\,mA. Please note that the dc sweeps and the two-dimensional maps were recorded on different samples. Due to minor differences in the lithographic patterning, there is slight mismatch in the auto-oscillation frequencies recorded at 4\,mA.} 
	\end{center}
\end{figure}


First, we saturated the structure at $\mu_{0} \mathrm{H}_\mathrm{ext}=500$\,mT and then reduced the value to $\mu_{0} \mathrm{H}_\mathrm{ext}=86$\,mT. In Fig.~\ref{fig2}a,b, we plot the simulated $x$- and $y$-component of the magnetisation in the resulting equilibrium state. As can be seen from the M$_{y}$-component (Fig.~\ref{fig2}b), the magnetisation in the centre of the nanowire width aligns along the external field.  Only at the boundaries, the magnetic moments line up with the edges to reduce the demagnetisation field, as evidenced by the non-vanishing M$_{x}$-component in Fig.~\ref{fig2}a. This state results in $\vec{\bf M} \bot \vec{\bf J}_\mathrm{c}$ only inside the 90$^{\circ}$ bends at positions A and B. In the straight parts of the nanowire, $ \vec{\bf J}_\mathrm{c}$ flows at a  $45^{\circ}$ angle with respect to $ \vec{\bf M}$, which strongly reduces the efficiency of the STT and, thus, the excitation of auto-oscillations. 

To observe these auto-oscillations, we use space-resolved Brillouin light scattering (BLS) microscopy \cite{Sebastian2015}. In Fig.~\ref{fig2}c,d, we plot the BLS intensity that was measured at positions A and B, respectively, as a function of the BLS frequency and the applied charge current which was swept consecutively from $-0.1$ to $-5$\,mA and from $0.1$ to $5$\,mA. For negative currents, no auto-oscillations are excited because $ \vec{\bf P}\|  \vec{\bf M}$ which leads to an increased effective damping. For positive currents, however, $ \vec{\bf P}$ is antiparallel to $ \vec{\bf M}$,  and at $ I_\mathrm{c}>2.2$\,mA one auto-oscillation mode is detected at a frequency of about 9\,GHz. For $ I_\mathrm{c}>3.4$\,mA, even two modes are observed, starting at 9.6\,GHz and 10.2\,GHz. All modes show a negative frequency shift with increasing  $  I_\mathrm{c}$ which is expected for in-plane magnetised SHNOs \cite{Slavin}.

In order to illustrate the spatial character of the excited auto-oscillations, we picked a direct current of $ I_\mathrm{c}=4$\,mA at $\mu_{0} \mathrm{H}_\mathrm{ext}=86$\,mT. At this specific current, two auto-oscillations are excited at 9.3 and 9.9\,GHz. The spatial intensity distributions in Fig.~\ref{fig2}e,f show a  strong localisation of the 9.3\,GHz mode at the apex whereas the mode at 9.9\,GHz is much more diverged. 
The localisation is related to the changing angle between $ \vec{\bf J}_\mathrm{c}$ and $\vec{\bf M}$ as well as the slightly increased current density at the 90$^{\circ}$ bends.

\begin{figure}
\begin{center}
\scalebox{1}{\includegraphics[width=8.3cm, clip]{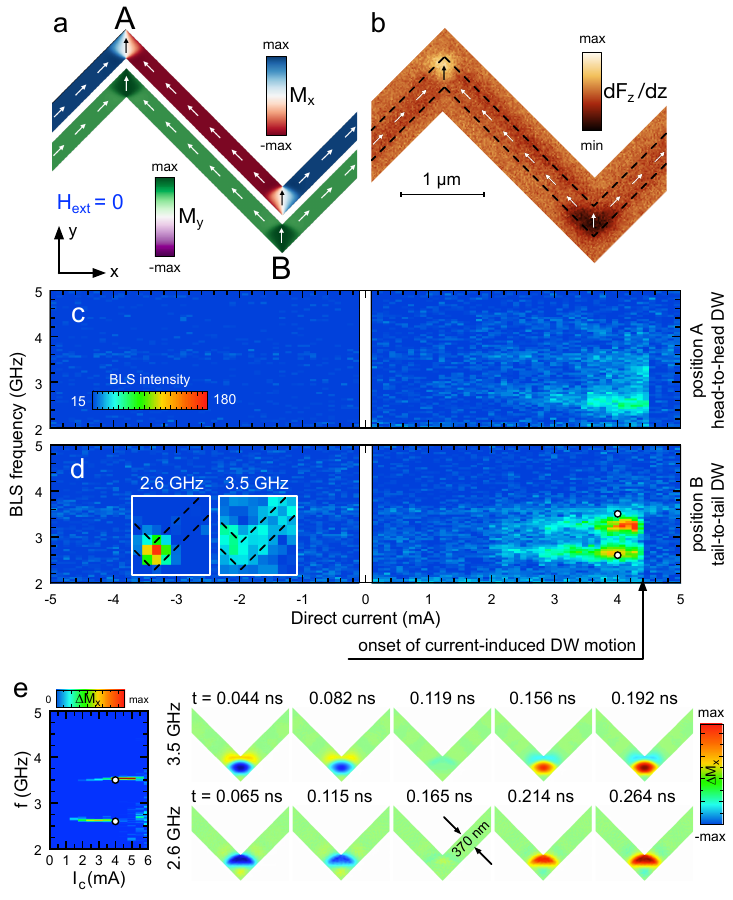}}
\caption{\label{fig3}\textbf{Auto-oscillations inside magnetic domain walls.} \textbf{a}, Micromagnetic simulation and \textbf{b}, MFM measurement of the remanent magnetisation after saturating the sample with $\mu_{0} \mathrm{H}_\mathrm{ext}=500$\,mT and reducing the field back to zero. {\bf c},{\bf d}, BLS spectra measured as a function of the applied $ I_\mathrm{c}$ on the DWs at positions A and B, respectively. The direct current was swept from $-0.1$ to $-5$\,mA and from $0.1$ to $5$\,mA, respectively. In between the consecutive current sweeps, the sample was saturated to ensure the proper domain configuration. Insets in {\bf d} show the two-dimensional BLS intensity distributions of auto-oscillations excited in the remanent state at 2.6 and 3.5\,GHz for $ I_\mathrm{c}=$\,4 \,mA. \textbf{e}, Micromagnetic simulations of the current sweep (left) and the temporal evolution (right) of the amplitudes of the simulated auto-oscillation modes excited at 2.6 and 3.5\,GHz for a half cycle with $ {I}_\mathrm{c}=4$\,mA.}
\end{center}
\end{figure}


Up to now, we always applied an external magnetic field in order to drive auto-oscillations within the nanowire. In the next step, we prepared a magnetic transverse DW at the 90° bends. Therefore, we again saturated the nanowire at $\mu_{0} \mathrm{H}_\mathrm{ext}=500$\,mT but then reduced the external magnetic field  to zero. Figure~\ref{fig3}a  depicts the resulting remanent state of the nanowire as derived from micromagnetic simulations. The shape anisotropy directs the magnetisation to align along the nanowire, which creates a head-to-head (tail-to-tail) DW at position A (B), respectively.

The domain configuration in the remanent state is also confirmed by  magnetic force microscopy (MFM) in Fig.~\ref{fig3}b. The bright (dark) contrast at position A (B)  in the dF$_{z}$/dz-signal confirms the formation of DWs where F$_{z}$ is the $z$ component of the magnetostatic tip-sample interaction. 

In the straight parts of the nanowire, the charge current flows along the magnetisation ($ \vec{\bf J}_\mathrm{c}\| \vec{\bf M}$) so that no auto-oscillations can be excited. Only inside the DWs at positions A and B, the direct current and the magnetisation are aligned perpendicularly so that we  expect to observe auto-oscillations only in these narrow regions. 

When we apply currents between $-0.1$ and $-5$\,mA and measure the BLS intensities inside the DWs at positions A and B (see Fig.~\ref{fig3}c,d respectively), we again see no auto-oscillations due to increased damping in this direction. For positive currents, however, we observe two auto-oscillation modes at $2.6$ and $3.5$\,GHz. These modes do not show a pronounced negative frequency shift and one of them is in fact strongly localised to the DW, as is confirmed by spatially resolved BLS microscopy (see insets in Fig.~\ref{fig3}d). The experimental results are further corroborated by micromagnetic simulations in Fig.~\ref{fig3}e, which show the localisation of both modes and a node in the dynamic magnetisation profile of the higher frequency mode. 

We would like to stress the fact that for $ I_\mathrm{c}>4.5$\,mA the measured BLS intensity at positions A and B abruptly disappears. Indeed, this is still the case when we decrease the current again. Only a repeated saturation and relaxation of the sample, as described before, restores the auto-oscillations. This illustrates that for $ I_\mathrm{c}>4.5$\,mA the DWs are removed from the apexes due to the current-induced DW motion.  
For lower currents, this motion is suppressed by the dipolar pinning of the DWs at the apexes. One can exploit this motion with a precise current pulse, e.\,g., to shift one DW from position A to position B.

In conclusion, we have demonstrated that a pinned transversal DW is able to work as a nano-sized spin-torque oscillator. Once prepared, this oscillator is able to perform at zero external magnetic field. Moreover, the auto-oscillations appear at current densities at which the current-induced DW motion not yet sets in. In other words, the DW oscillates at GHz frequencies before it moves. In principle, the ratio of these two thresholds can be tuned by changing the relative thicknesses or materials of the functional layers. If desired, the DW motion could also be completely shut off by replacing the metallic ferromagnet with an insulating one. On the other hand, the combination of auto-oscillations and current-induced DW motion yields various possibilities for research in fundamental DW physics as well as for applications. One could think of using moveable DW spin-torque oscillators in networks to achieve more flexibility in synchronisation or frequency locking or even further advance concepts for magnetic racetrack memories or neuromorphic computing.

\section*{Methods}
{\bf Sample Preparation.} A zigzag structure was fabricated from Cr(1.5)/Pt(7)/Co$_{40}$Fe$_{40}$B$_{20}$(5)/Cr(1.5) multilayer on Si/SiO$_{2}$ substrate using electron beam lithography, magnetron sputtering and lift off. On the zigzag structure, a pair of electric contacts was made from a Ta(5)/Cu(150)/Ta(5) multilayer to inject a direct current to the Pt layer. All thicknesses are given in nanometres.

{\bf Magnetic force microscopy.}
In order to prepare the remanent state, an external field of $\mu_{0} \mathrm{H}_\mathrm{ext}=500$\,mT was applied and slowly reduced to zero before starting  MFM measurements. The measurements were performed in a Nanoscan high resolution magnetic force microscope in high vacuum using a high aspect ratio MFM probe (Team Nanotec HR-MFM45 with ML1 coating and a spring constant of 0.7\,N/m). The local tip-sample distance was 82\,nm.

{\bf Micromagnetic simulation} The micromagnetic ground states as well as the auto-oscillation modes were obtained using the code from the GPU-based finite-difference micromagnetic package MuMax3.\cite{mumax}
The material parameters used in the simulations include the saturation magnetization $\mu_{0}M_{s}=1.28$\,T, 
exchange stiffness of $A_{\mathrm{ex}}=20$\,pJ/m, and damping parameter $\alpha=0.001$. We ran the simulations for a geometry with a lateral size of 4000$\times$3000$\times$5\,nm$^3$, and unit cell size 3.90625$\times$3.90625$\times$5\,nm$^3$. Absorbing boundary conditions in the form of a smooth increase of the damping profile are applied to the open edges of the nanowire to avoid artifacts from  spin-wave  reflection. The direct charge current transport were simulated using COMSOL Multiphysics (\href{www.comsol.com}{www.comsol.com}) using layer resistivities of 90\,$\mu\Omega$cm and 11.2\,$\mu\Omega$cm for CoFeB and Pt, respectively. The current profile and the Oersted field are then supplied to MuMax3, taking into account a spin-Hall angle of $\mathit{\theta_{SH}}$=0.08 and the calculated spatial distribution of the spin current polarization. The auto-oscillation spectra were obtained by applying a Fast Fourier Transform (FFT) to the net $x$ component of the total magnetization, simulated over 100\,ns. The spatial profiles of the auto-oscillation modes are constructed via an FFT for each simulation cell \cite{McMichael2005}.
The auto-oscillation modes show very similar profiles and frequencies as the linear modes excited with a field pulse at zero current, in other words the DW auto-oscillations nucleate from linear modes, similar to the high frequency modes in other SHNOs \cite{DvornikPRA2018}.

{\bf Brillouin light scattering microscopy.} All measurements were performed at room temperature. The auto-oscillation intensity is locally recorded by means of BLS microscopy. This method is based on the inelastic scattering of light and magnetic oscillations. Light from a continuous wave, single-frequency 532-nm solid-state laser is focused on the sample surface using a high numerical aperture microscope lens giving a spatial resolution of 340\,nm. The laser power on the sample surface is typically about 1\,mW. The frequency shift of the inelastically scattered light is analysed using a six-pass Fabry-Perot interferometer TFP-2 (JRS Scientific Instruments). To record two-dimensional maps of the intensity distribution, the sample is moved via a high-precision translation stage (10-nm step size, Newport). The sample position is continuously monitored with a CCD camera using the same microscope lens. A home-built active stabilisation algorithm based on picture recognition allows for controlling the sample position with respect to the laser focus with a precision better than 20\,nm.

\section*{acknowledgments}
Financial support by the Deutsche Forschungsgemeinschaft within programme SCHU2922/1-1 is gratefully acknowledged. N.S. acknowledges funding from the Alexander von Humboldt Foundation. K.S. acknowledges funding from the Helmholtz Postdoc Programme. Samples were fabricated at the Nanofabrication Facilities (NanoFaRo) at the Institute of Ion Beam Physics and Materials Research at HZDR. We thank Dr. Ingolf M\"onch for deposition of the Ta/Cu/Ta multilayer. 

\section*{Author contributions}
H.S., N.S. conceived and designed the experiments. 
N.S. and S.S.P.K.A performed lithographic processing and fabricated the thin film.
L.K. performed static micromagnetic simulations, A.A.A. performed  COMSOL modeling and dynamical micromagnetic simulations.
N.S. performed BLS experiments. 
N.P. and T.M. performed MFM measurements. 
H.S., K.S. and L.K. wrote the manuscript. 
All authors discussed the results and commented on the manuscript.

\section*{Additional information}
The authors declare no competing financial interests. Reprints and permission information is available online at http://www.nature.com/reprints. 

\section*{References}

\begin{thebibliography}{99}

\bibitem{Parkin2008} Parkin, S.~S.~P., Hayashi, M. \& Thomas, L. Magnetic domain-wall racetrack memory. {\it Science} {\bf 320}, 190-194 (2008).
\bibitem{Locatelli} Locatelli, N., Cros, V. \& Grollier, J. Spin-torque building blocks. {\it Nat. Publ. Group} {\bf 13}, 11-20 (2014).
\bibitem{Grollier2017} Torrejon, J. {\it et al.} Neuromorphic computing with nanoscale spintronic oscillators. {\it Nature} {\bf 547}, 428-431 (2017).
\bibitem{Slonczewski1996} Slonczewski, J.~C. Current-driven excitation of magnetic multilayers. {\it J. Magn. Magn. Mater.} {\bf 159}, L1L7 (1996).
\bibitem{Berger1996} Berger, L. Emission of spin waves by a magnetic multilayer traversed by a current. {\it Phys. Rev. B} {\bf 54}, 93539358 (1996).
\bibitem{Myers} Myers, E.  {\it et al.}  Current-induced switching of domains in magnetic multilayer devices. {\it Science} {\bf 285}, 867 (1999).
\bibitem{Kiselev} Kiselev, S.~I. {\it et al.} Microwave oscillations of a nanomagnet driven by a spin-polarized current {\it Nature} {\bf 425}, 380 (2003).
\bibitem{CIDWM} Beach, G., Tsoi, M., \& Erskine J.~L. Current-induced domain wall motion. {\it Journal of Magnetism and Magnetic Materials} {\bf 320}, 1272 (2008)
\bibitem{Skyrme}Skyrme, T.~H.~R. A unified field theory of mesons and baryons. {\it Nucl. Phys.} {\bf 31}, 556569 (1962).
\bibitem{Muhlbauer} M\"uhlbauer, S. {\it et al.} Skyrmion lattice in a chiral magnet. {\it Science} {\bf 323}, 915919 (2009).
\bibitem{Fert2013} Fert, A., Cros, V. \& Sampaio, J. Skyrmions on the track. {\it Nature Nano.} {\bf 8}, 152-156 (2013).
\bibitem{Slavin} Slavin, A.~N. \& Tiberkevich, V. Nonlinear auto-oscillator theory of microwave generation by spin-polarized current, \textit{IEEE Transactions on Magnetics}, {\bf 45}, 1875 (2009).
\bibitem{Madami} Madami, M. {\it et al.}, Direct observation of a propagating spin wave induced by spin-transfer torque. \textit{Nature Nanotech.} 6, 635 (2011).
\bibitem{Dyakonov1971} Dyakonov, M. I. \& Perel, V. I. Possibility of orienting electron spins with current. {\it Sov. Phys. JETP Lett.} {\bf 13}, 467469 (1971). 
\bibitem{Hirsch1999}Hirsch, J.~E. Spin Hall effect. {\it Phys. Rev. Lett.} {\bf 83}, 18341837 (1999). 
\bibitem{Ando2008} Ando, K. {\it et al.} Electric manipulation of spin relaxation using the spin Hall effect. {\it Phys. Rev. Lett.} {\bf 101}, 036601 (2008).
\bibitem{Hoffmann2013}Hoffmann, A. {\it IEEE Trans. Magn.}  {\bf 49}, 5172 (2013).
\bibitem{Urazhdin2016} Urazhdin, S., {\it et al.} Excitation of coherent propagating spin waves by pure spin currents. {\it Nat. Commun.} {\bf 7}, 1-6 (2016).
\bibitem{Divinskiy2018} Divinskiy, B. {\it et al.} Excitation and amplification of spin waves by spin-orbit torque. \textit{Advanced Materials} {\textbf 30}, 1802837 (2018).
\bibitem{Neusser2009}Neusser, S. \& Grundler, D. Magnonics: spin waves on the nanoscale. {\it Adv. Mater.} {\bf 21}, 29272932 (2009).
\bibitem{Kruglyak2010}Kruglyak, V.~V., Demokritov, S.~O. \& Grundler, D. Magnonics. {\it J. Phys. D} {\bf 43}, 264001 (2010).
\bibitem{Lenk2011}Lenk, B., Ulrichs, H., Garbs, F. \& M\"unzenberg, M. The building blocks of magnonics. {\it Phys. Rep.} {\bf 507}, 107136 (2011).
\bibitem{Chumak2015} Chumak, A.~V., Vasyuchka, V.~I., Serga, A.~A. \& Hillbrands, B. Magnon spintronics. {\it Nature Physics } {\bf 11}, 453461 (2015).
\bibitem{Wagner} Wagner, K. {\it et al.} Magnetic domain walls as reconfigurable spin-wave nanochannels. {\it Nature Nanotech.} {\bf 11}, 432 (2016).
\bibitem{Duan} Duan, Z., {\it et al.} Nanowire spin torque oscillator driven by spin orbit torques. {\it Nat. Commun.} {\bf 5}, 5616 (2014).
\bibitem{Demidov2012} Demidov, V.~E.,  {\it et al.} Magnetic nano-oscillator driven by pure spin current. {\it Nature Materials} {\bf 11}, 1028-1031 (2012).
\bibitem{Demidov2014} Demidov, V.~E., Urazhdin, S.,  Zholud, A.,  Sadovnikov,  A.~V., \&  Demokritov, S.~O. Nanoconstriction-based spin-Hall nano-oscillator  {\it Appl. Phys. Lett.} {\bf 105}, 172410 (2014).
\bibitem{Amad}D\"urrenfeld, P.,  Awad, A.~A., Houshang, A., Dumas, R.~K., \& \AA kerman, J.~A., 20 nm spin Hall nano-oscillator {\it Nanoscale} {\bf9}, 1285 (2017).
\bibitem{Sebastian2015}Sebastian, T., Schultheiss, K., Obry, B., Hillebrands, B. \& Schultheiss, H. Micro-focused Brillouin light scattering: imaging spin waves at the nanoscale.  {\it  Front. Phys}. {\bf 3}, 35 (2015).

\bibitem{mumax} Vansteenkiste, A. {\it et al.} The design and verification of Mumax3. {\it  AIP Advances}. {\bf 10}, 107133 (2014).
\bibitem{McMichael2005} McMichael, R.~D. \& Stiles, M.~D., Magnetic normal modes of nanoelements. {\it  J. Appl. Phys.}. {\bf 10}, 10J901 (2005).
\bibitem{DvornikPRA2018} Dvornik, M., Awad, A.~A. \& \AA{}kerman, J., Origin of Magnetization Auto-Oscillations in Constriction-Based Spin Hall Nano-Oscillators. {\it  Phys. Rev. Applied}. {\bf 9}, 014017 (2018).


\end{thebibliography}
\end{document}